\documentclass[
aps,nature,showpacs,superscriptaddress,numerical,floatfix,
amsmath,amssymb,
reprint,%
]{revtex4-1}

\usepackage{amsmath,amssymb,bm}
\usepackage[overload]{empheq}
\usepackage{mathtools}
\usepackage{xcolor}
\usepackage[
	pdffitwindow=true,
	colorlinks=true,
	frenchlinks=false,
        linkcolor=blue,
	anchorcolor=blue,
        citecolor=blue,
        filecolor=blue,
        urlcolor=blue,
        bookmarks=true,
        bookmarksopen=true,
	bookmarksnumbered=true,
        bookmarksopenlevel=1,
        plainpages=false,
	pdfpagelayout=TwoPageLeft,
        pdfpagelabels=true,
	breaklinks
]{hyperref}
\usepackage[per-mode=symbol,separate-uncertainty]{siunitx}
\usepackage{graphicx}
\usepackage{dcolumn}
\usepackage{bm}
\usepackage[english]{babel}
\usepackage{color}
\usepackage{nicefrac}

\begin{document}
\preprint{AIP/123-QED}
\title[]{Low Temperature Suppression of the Spin Nernst Angle in Pt}

\author{T.~Wimmer}
\email[]{tobias.wimmer@wmi.badw.de}
\affiliation{Walther-Mei{\ss}ner-Institut, Bayerische Akademie der Wissenschaften, 85748 Garching, Germany}
\affiliation{Physik-Department, Technische Universit\"{a}t M\"{u}nchen, 85748 Garching, Germany}
\author{J.~G{\"u}ckelhorn}
\affiliation{Walther-Mei{\ss}ner-Institut, Bayerische Akademie der Wissenschaften, 85748 Garching, Germany}
\affiliation{Physik-Department, Technische Universit\"{a}t M\"{u}nchen, 85748 Garching, Germany}
\author{S.~Wimmer}
\affiliation{Department Chemie, Physikalische Chemie, Universit\"{a}t M\"{u}nchen, Butenandtstraße 5-13, 81377 M\"{u}nchen, Germany}
\author{S.~Mankovsky}
\affiliation{Department Chemie, Physikalische Chemie, Universit\"{a}t M\"{u}nchen, Butenandtstraße 5-13, 81377 M\"{u}nchen, Germany}
\author{H.~Ebert}
\affiliation{Department Chemie, Physikalische Chemie, Universit\"{a}t M\"{u}nchen, Butenandtstraße 5-13, 81377 M\"{u}nchen, Germany}
\author{M.~Opel}
\affiliation{Walther-Mei{\ss}ner-Institut, Bayerische Akademie der Wissenschaften, 85748 Garching, Germany}
\author{S.~Gepr{\"a}gs}
\affiliation{Walther-Mei{\ss}ner-Institut, Bayerische Akademie der Wissenschaften, 85748 Garching, Germany}
\author{R.~Gross}
\affiliation{Walther-Mei{\ss}ner-Institut, Bayerische Akademie der Wissenschaften, 85748 Garching, Germany}
\affiliation{Physik-Department, Technische Universit\"{a}t M\"{u}nchen, 85748 Garching, Germany}
\affiliation{Munich Center for Quantum Science and Technology (MCQST), Schellingstr. 4, D-80799 M\"{u}nchen, Germany}
\author{H.~Huebl}
\affiliation{Walther-Mei{\ss}ner-Institut, Bayerische Akademie der Wissenschaften, 85748 Garching, Germany}
\affiliation{Physik-Department, Technische Universit\"{a}t M\"{u}nchen, 85748 Garching, Germany}
\affiliation{Munich Center for Quantum Science and Technology (MCQST), Schellingstr. 4, D-80799 M\"{u}nchen, Germany}
\author{M.~Althammer}
\email[]{matthias.althammer@wmi.badw.de}
\affiliation{Walther-Mei{\ss}ner-Institut, Bayerische Akademie der Wissenschaften, 85748 Garching, Germany}
\affiliation{Physik-Department, Technische Universit\"{a}t M\"{u}nchen, 85748 Garching, Germany}

\date{\today}


\begin{abstract}
We demonstrate the low temperature suppression of the platinum (Pt) spin Nernst angle in bilayers consisting of the antiferromagnetic insulator hematite ($\alpha$-Fe$_2$O$_3$) and Pt upon measuring the transverse spin Nernst magnetothermopower (TSNM). We show that the observed signal stems from the interplay between the interfacial spin accumulation in Pt originating from the spin Nernst effect and the orientation of the N\'eel vector of $\alpha$-Fe$_2$O$_3$, rather than its net magnetization. Since the latter is negligible in an antiferromagnet, our device is superior to ferromagnetic structures, allowing to unambiguously distinguish the TSNM from thermally excited magnon transport (TMT), which usually dominates in ferri/ferromagnets due to their non-zero magnetization. Evaluating the temperature dependence of the effect, we observe a vanishing TSNM below $\sim\SI{100}{\kelvin}$. We compare these results with theoretical calculations of the temperature dependent spin Nernst conductivity and find excellent agreement. This provides evidence for a vanishing spin Nernst angle of Pt at low temperatures and the dominance of extrinsic contributions to the spin Nernst effect. 
\end{abstract}


\maketitle

The observation of the spin Nernst effect (SNE) in 2017~\cite{Meyer2017,Sheng2017,Kim2017,Bose2018} has, together with the spin Hall (SHE)~\cite{KatoSHE,Wunderlich2005,Valenzuela2006,SaitohISHE,Wunderlich2010,Sinova2015}, the spin Seebeck~\cite{Uchida2008,Jaworski2010,Uchida2010} and the spin Peltier effect~\cite{Flipse2012,Flipse2014}, completed the picture of electronic transport phenomena based on the coupling of charge, heat and spin transport. A typical way to detect these spin transport phenomena is to modify the boundary conditions with bilayer structures consisting of a heavy metal (HM) and a magnetically ordered insulator (MOI)~\cite{NakayamaSMR,AltiSMR,ChenSMR}, where the relative orientation of the spin accumulation vector $\bm{s}$ in the HM and the (sublattice) magnetization vector $\bm{M}$ of the MOI at their shared interface is of key importance. It determines the spin current transmission through the interface, leading to a magnetization direction dependent magnetoresistance/magnetothermopower effect.

Recently, the interplay of pure spin currents generated in HM films with antiferromagnetic insulators (AFIs) has gained great attention~\cite{Klaui2018,Opel2018,Opel2020,Wimmer2020}. While this has been investigated by the spin Hall magnetoresistance (SMR)~\cite{Hoogeboom2017,Opel2018,Opel2020} and long distance magnon transport~\cite{Klaui2018,Han2020,Wimmer2020}, the interplay of spin currents generated via the SNE with adjacent AFIs has not yet been studied.

In this Letter, we report on the low temperature suppression of the Pt spin Nernst angle. For this purpose, we measure the temperature dependence of the TSNM in a HM Pt thin film deposited on the antiferromagnetic insulator hematite ($\alpha$-Fe$_2$O$_3$). The use of an antiferromagnet is beneficial, since it allows to clearly distinguish between the SNE and other thermopower effects that are related to the small net magnetization. We explain our results by a decreasing spin Nernst conductivity with decreasing temperature calculated from a first principles spin transport theory~\cite{SWimmer2013,Meyer2017}. This suggests a vanishing spin Nernst angle of Pt below $\SI{100}{\kelvin}$ and highlights the importance of extrinsic contributions to the SNE and SHE in our Pt layers. 

\begin{figure}[]%
	\includegraphics[]{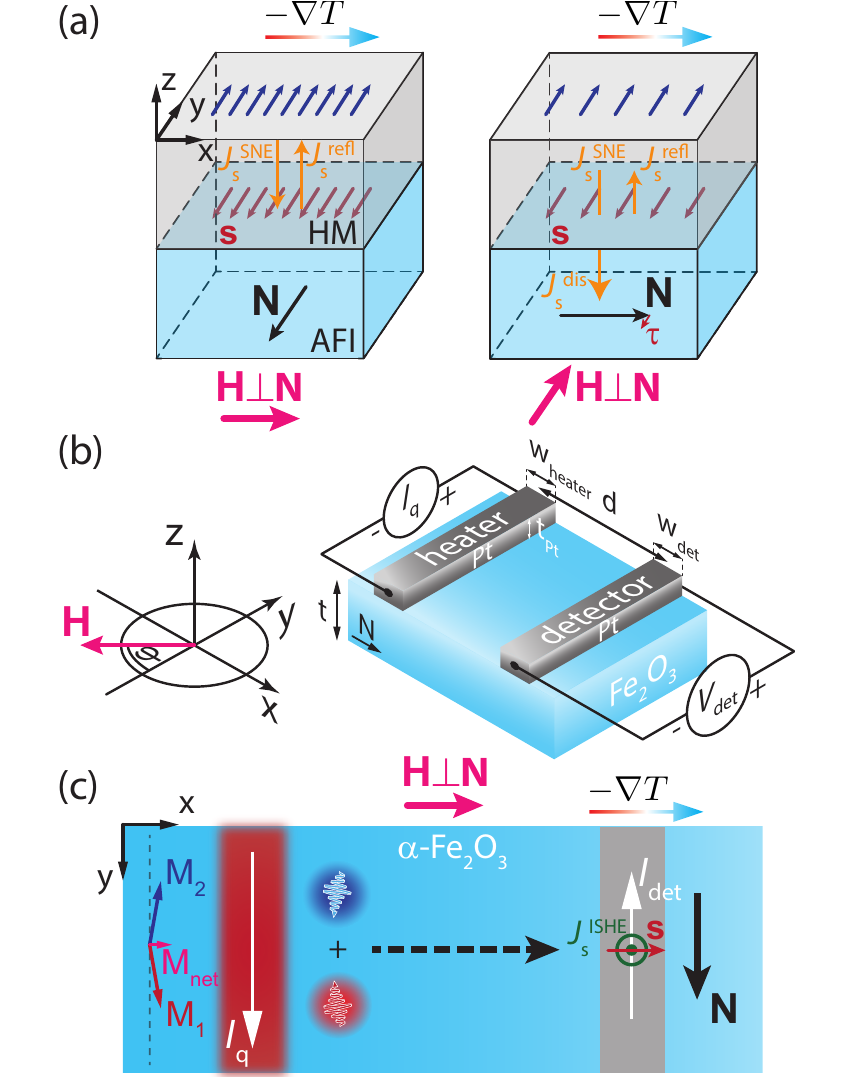}%
	\caption{(a) Sketch of an AFI/HM bilayer. In open circuit conditions, a temperature gradient generates a spin accumulation $\bm{s}$ at the AFI/Pt interface with $\bm{s} \perp -\nabla T$. Depending on the relative orientation of $\bm{s}$ and $\bm{N}$, the spin accumulation at the interface is either unaffected by the AFI (for $\bm{s}\parallel \bm{N}$, left panel) or partially dissipated via a spin current $J_\mathrm{s}^\mathrm{dis}$ in the AFI (for $\bm{s}\perp \bm{N}$, right panel). (b) Schematic depiction of the device, the electrical connection scheme and the coordinate system. A charge current $I_\mathrm{q}$ is fed through the left Pt electrode acting as a heat source (heater). To determine the transverse magnetothermopower, a voltage signal $V_\mathrm{det}$ in a second Pt-stripe (detector) is measured as a function of the magnetic field orientation $\varphi$. (c) Joule heating induces a thermal injection of the two antiferromagnetic magnon modes (blue and red wiggly arrows). The effective magnetic moment is given by their superposition and is dominantly proportional to the canted net magnetization $\bm{M}_\mathrm{net}$. The diffusing magnons are measured at the Pt detector via the inverse SHE. Simultaneously, a lateral temperature gradient across the width of the Pt detector leads to the emergence of the TSNM. The canting of the two sublattice magnetizations $\bm{M}_{1,2}$ in $\alpha$-Fe$_2$O$_3$ leading to a net magnetization $\bm{M}_{net}$ is schematically depicted on the left.
	}%
	\label{fig:scheme}%
\end{figure}

The principle of the spin Nernst magnetothermopower in an AFI/HM bilayer is depicted in Fig.~\ref{fig:scheme}(a). 
In open circuit conditions, neither charge nor spin currents can flow in the HM. Applying a temperature gradient $-\nabla T$ along the $\bm{x}$-direction leads to the generation of an electrochemical and spin-chemical potential (i.e.~spin accumulation) counteracting the conventional Seebeck and spin Nernst current via the emerging diffusive charge/spin currents. Consequently, the spin accumulation $\bm{s}$ in the HM builds up at the top and bottom interface along $\bm{y}$. 
Its relative orientation to the Néel vector $\bm{N}$ then determines the relevant boundary condition: for $\bm{N}\parallel\bm{s}$ (left panel in Fig.~\ref{fig:scheme}(a)), the spin accumulation cannot dissipate in the AFI. If $\bm{s}\nparallel\bm{N}$ (right panel in Fig.~\ref{fig:scheme}(a)), a spin transfer torque $\bm{\tau}$ can be exerted on $\bm{N}$~\cite{Cheng2015}, leading to a finite spin current $J_\mathrm{s}^\mathrm{dis}$ dissipating in the AFI and a reduction of $J_\mathrm{s}^\mathrm{SNE}$ in the HM layer. While $J_\mathrm{s}^\mathrm{dis}$ decays within the antiferromagnetic hematite, the reflected spin current $J_\mathrm{s}^\mathrm{refl}$ is converted back to a charge current via the inverse SHE. In open circuit conditions, an electric field across the $x$-$y$-plane of the HM arises, the direction of which depends upon the relative orientation of $\bm{N}$ to $\bm{s}$.

In our experiment, we employ an epitaxial (0001)-oriented antiferromagnetic hematite thin film with a thickness of $t=\SI{15}{\nano\metre}$ grown on a sapphire (Al$_2$O$_3$) substrate. Details of the growth process are published elsewhere~\cite{Wimmer2020}.~The magnetic phase of our hematite films features an easy-plane anisotropy with two antiferromagnetically coupled sublattice magnetizations $\bm{M}_1$ and $\bm{M}_2$, exhibiting a slight canting due to a crystalline Dzyaloshinskii-Moriya interaction (DMI)~\cite{Dzyaloshinsky1958,Moriya1960}. 
As a consequence, a net magnetization $\bm{M}_\mathrm{net} = \bm{M}_1 + \bm{M}_2$ exists even for zero external magnetic field strength $\mu_0 H$. As $\mu_0 \bm{H}$ couples to $\bm{M}_\mathrm{net}$, the antiferromagnetic N\'eel vector $\bm{N} = (\bm{M}_1 - \bm{M}_2$)/2 satisfies $\bm{N} \perp \bm{H}$ as long as the external field exceeds the magnetic field for a single domain state given by $\mu_0 H_\mathrm{SD}\approx\SI{3}{\tesla}$ in our $\alpha$-Fe$_2$O$_3$/Pt bilayers~\cite{Gepraegs2020}. Below $\mu_0 H_\mathrm{SD}$, the threefold crystalline anisotropy leads to the emergence of magnetic $120^\circ$ domains, which are equally distributed for $\mu_0 H = 0$~\cite{Marmeggi1977}.

Pt electrodes with thickness $t_\mathrm{Pt} = \SI{5}{\nano\metre}$ are deposited on the hematite thin film via sputter deposition and patterned into multiple pairs of nanostrips via electron beam lithography and lift-off. A sketch of a typical device is given in Fig.~\ref{fig:scheme}(b) (see footnote \footnote{After the Pt deposition, we use $\SI{50}{\nano\metre}$ of aluminum (Al) sputtered on the film which is patterned into leads and bonding pads to electrically connect the Pt electrodes. The lengths of the heater and detector strips are $\SI{162}{\micro\metre}$ and $\SI{148}{\micro\metre}$, respectively. Three different devices with heater strip widths $w_\mathrm{heater} = \SI{200}{\nano\metre}$, $\SI{500}{\nano\metre}$, $\SI{600}{\nano\metre}$ are investigated. The detector width is constant with $w_\mathrm{det} = \SI{500}{\nano\metre}$ for all three devices.} for geometrical details). 
The distance between the strips is characterized by the center-to-center strip separation $d$. A charge current $I_\mathrm{q}$ corresponding to a current density $J_\mathrm{q} \sim \SI{2e11}{\ampere\per\metre\squared}$ is applied to the left Pt (heater) strip and used for local Joule heating. The resulting voltage signal $V_\mathrm{det}$ along the length of the right Pt (detector) strip is measured as a function of the in-plane orientation $\varphi$ of the external magnetic field (c.f.~Fig.~\ref{fig:scheme}(b)). We extract contributions of thermal origin via the current reversal method~\cite{Gueckelhorn2020,SchlitzMMR,KathrinLogik} by investigating the signal $V_\mathrm{det}^\mathrm{th} = [V_\mathrm{det}(+I_\mathrm{q}) + V_\mathrm{det}(-I_\mathrm{q})]/2$.

As a consequence of the local heating, two thermally-induced effects are present at the detector. The first effect and main focus of this work refers to the lateral temperature gradient that emerges across the width of the Pt detector electrode depicted in Fig.~\ref{fig:scheme}(c). As discussed in the context of Fig.~\ref{fig:scheme}(a), this configuration gives rise to a magnetothermopower induced by the SNE. In our particular experimental configuration, we measure its transverse contribution, which we denote as the TSNM. Based on the change of the boundary condition of the SNE-induced spin accumulation at the interface with the orientation of the applied magnetic field, its conversion to a voltage signal is expected to follow an angle dependence proportional to $\sin(2\varphi)$. For $\bm{N}\parallel\bm{s}$ (i.e.~$\varphi=90^\circ,270^\circ$), the SNE-induced spin accumulation is unaffected by the magnetic order $\bm{N}$ in the AFI. Since we measure the transverse voltage drop along the length of the Pt detector ($\bm{y}$-direction), the maximum voltage signals are expected for $\varphi = 90^\circ \pm 45^\circ,270^\circ \pm 45^\circ$ and are therefore shifted by $45^\circ$ compared to longitudinal measurements~\cite{Meyer2017}. Hence, we expect $V_\mathrm{det}^\mathrm{th}\propto\sin(\varphi)\cos(\varphi) \propto \sin(2\varphi)$. 

The second effect, as depicted in Fig.~\ref{fig:scheme}(c), refers to a thermal injection of the antiferromagnetic magnon modes leading to a locally excited non-equilibrium magnon distribution that diffuses throughout the antiferromagnet~\cite{Klaui2018}. 
Since the thermal magnon injection is not sensitive to the spin polarization direction (unlike the injection of spin current via the SHE~\cite{Klaui2018,Wimmer2020} and SNE), both antiferromagnetic magnon modes with opposite chirality (i.e.~spin polarization) are simultaneously excited (see the blue and red wiggly arrows in Fig.~\ref{fig:scheme}(c)). Thus, the transported spin is given by a superposition of these two excitations. Since the frequencies of the two modes are generally non-degenerate for a canted antiferromagnetic state~\cite{Rezende2019}, we expect spin transport contributions polarized along both the N\'{e}el vector $\bm{N}$ as well as the net magnetization $\bm{M}_\mathrm{net}$. However, as reported in Ref.~\onlinecite{Klaui2018} as well as supported by our data, the contribution from $\bm{N}$ to the thermally excited magnon transport effect (TMT) turns out to be negligible for the typical length scales investigated. We can therefore assume that the TMT is proportional to the net magnetization $\bm{M}_\mathrm{net}$~\cite{Klaui2018}. At the detector, the diffusing magnon accumulation is converted into a charge current via the ISHE. Due to its symmetry, the angle dependence of the corresponding detector signal follows $V_\mathrm{det}^\mathrm{th} \propto \sin(\varphi)$~\cite{Klaui2018}.

\begin{figure}[htb!]%
	\includegraphics[]{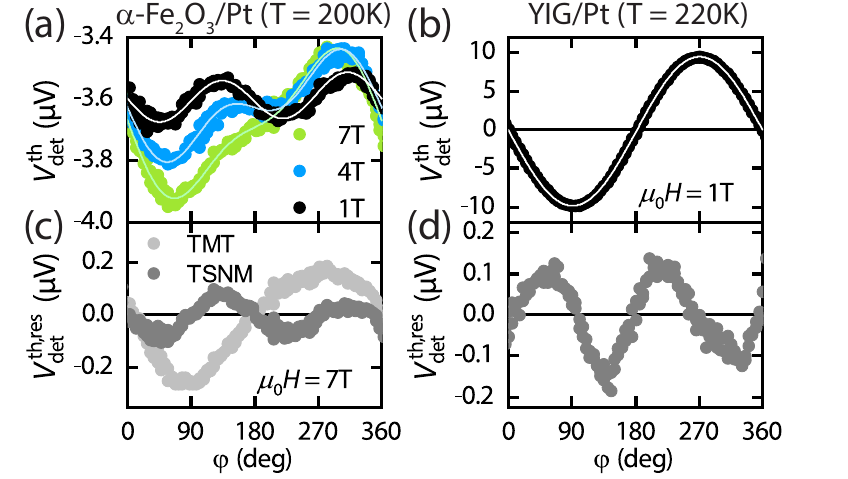}%
	\caption{(a) Detector signal $V_\mathrm{det}^\mathrm{th}$ as a function of magnetic field orientation $\varphi$ for different external magnetic field strengths, where a superposition of both the TMT and TSNM signal is observed. The data is shown for a device with heater-detector distance $d=\SI{750}{\nano\metre}$ at $T=\SI{200}{\kelvin}$. Solid lines are fits to Eq.~\eqref{eq:fit}. (b) Detector signal $V_\mathrm{det}^\mathrm{th}$ measured on a similar device fabricated on the ferrimagnetic insulator Y$_3$Fe$_5$O$_{12}$ (YIG) for $d=\SI{2.1}{\micro\metre}$ at $T=\SI{220}{\kelvin}$ and $\mu_0 H = \SI{1}{\tesla}$. The white solid line is a fit to a $\sin(\varphi)$-type function. (c) Residual signals $V_\mathrm{det}^\mathrm{th,res}$ extracted from panel (a) using the different angular symmetries of the TMT (light gray points) and TSNM (dark gray points) for $\mu_0 H = \SI{7}{\tesla}$. (d) Residual signal $V_\mathrm{det}^\mathrm{th,res}$ of the fit to the data shown in panel (b), exhibiting the TSNM in YIG/Pt. A $90^\circ$ phase shift is observed compared to the TSNM in hematite. 
	}%
	\label{fig:ADMR}%
\end{figure}

The total angle dependence of the thermal detector signal $V_\mathrm{det}^\mathrm{th}$ can thus be expressed as

\begin{equation}
V_\mathrm{det}^\mathrm{th}(\varphi) = V_0 + \Delta V_\mathrm{det}^\mathrm{TMT}\sin{\varphi} + \Delta V_\mathrm{det}^\mathrm{TSNM}\sin{2\varphi},
\label{eq:fit}
\end{equation}
where $V_0$ is a constant offset voltage due to conventional thermal voltages independent of $\varphi$, $V_\mathrm{det}^\mathrm{TMT}$ is the amplitude of the TMT in the hematite and $V_\mathrm{det}^\mathrm{TSNM}$ represents the amplitude of the TSNM. Typical angle dependent measurements of $V_\mathrm{det}^\mathrm{th}$ at the detector are shown in Fig.~\ref{fig:ADMR}(a) for various external magnetic field strengths. The solid lines are fits to Eq.~\eqref{eq:fit}. Clearly, we find an excellent agreement of the fit with the experimental data. 
In order to separate the TSNM from the TMT, we extract the $180^\circ$-symmetric and $360^\circ$-symmetric signals stemming from the TSNM and TMT, respectively, which we denote as the residual signal $V_\mathrm{det}^\mathrm{th,res}$~\footnote{The TSNM and TMT signals are extracted by separately fitting the total signals shown in Fig.~\ref{fig:ADMR}(a) by a $\sin(\varphi)$-type and $\sin(2\varphi)$-type function, respectively, and plotting the residuals of the fits in panel (c).} (see Fig.~\ref{fig:ADMR}(c)) for $\mu_0 H = \SI{7}{\tesla}$. As expected, the $360^\circ$-symmetric modulation due to the TMT (light gray points) shows a minimum (maximum) signal for $\varphi = 90^\circ$ ($\varphi = 270^\circ$), where $\bm{M}_\mathrm{net}$ points perpendicular to the Pt detector. The $180^\circ$-symmetric signal due to the TSNM in Fig.~\ref{fig:ADMR}(c) (dark gray points) shows the expected $\sin(2\varphi)$ modulation. In order to experimentally demonstrate whether the TSNM is determined by the interaction of the spin polarization $\bm{s}$ with either the N\'eel vector $\bm{N}$ or the net magnetization $\bm{M}_\mathrm{net}$, we compare the results presented for hematite (Figs.~\ref{fig:ADMR}(a) and (c)) with a reference sample using the ferrimagnetic insulator yttrium iron garnet (Y$_3$Fe$_5$O$_{12}$, YIG). As shown in Fig.~\ref{fig:ADMR}(b), the thermal detector signal measured on the YIG sample shows a large $\sin(\varphi)$-type modulation due to the TMT~\cite{CornelissenMMR}, which is fitted to the data as the white solid line. In Fig.~\ref{fig:ADMR}(d), the residual of this fit is shown, demonstrating a clear $\sin(2\varphi)$ signature indicating the TSNM in YIG/Pt~\cite{Meyer2017}. Most interestingely, however, we observe a $90^\circ$ phase shift of the TSNM signal in hematite (Fig.~\ref{fig:ADMR}(c)) compared to YIG in Fig.~\ref{fig:ADMR}(d). Since $\bm{H}\perp\bm{N}$ in hematite and $\bm{H}\parallel\bm{M}_\mathrm{YIG}$ in YIG (with $\bm{M}_\mathrm{YIG}$ the YIG magnetization vector), we infer that the TSNM in the antiferromagnetic insulator hematite is indeed determined by $\bm{N}$ rather than $\bm{M}_\mathrm{net}$. This is consistent with the SMR effect in AFI/Pt bilayers~\cite{Opel2018,Opel2020,Gepraegs2020}. The comparison between the thermal detector signals in YIG/Pt and $\alpha$-Fe$_2$O$_3$/Pt highlights a further crucial difference: due to the much larger net magnetization $\bm{M}_\mathrm{YIG}$ as compared to the field-induced net magnetization $\bm{M}_\mathrm{net}$ of hematite, the TMT is blatantly dominant in YIG (c.f.~Fig.~\ref{fig:ADMR}(b)). In contrast, the N\'eel vector and the small net magnetization in hematite allow for an unambiguous, easily accessible differentiation of the TSNM and TMT, respectively.

\begin{figure}[b!]%
	\includegraphics[]{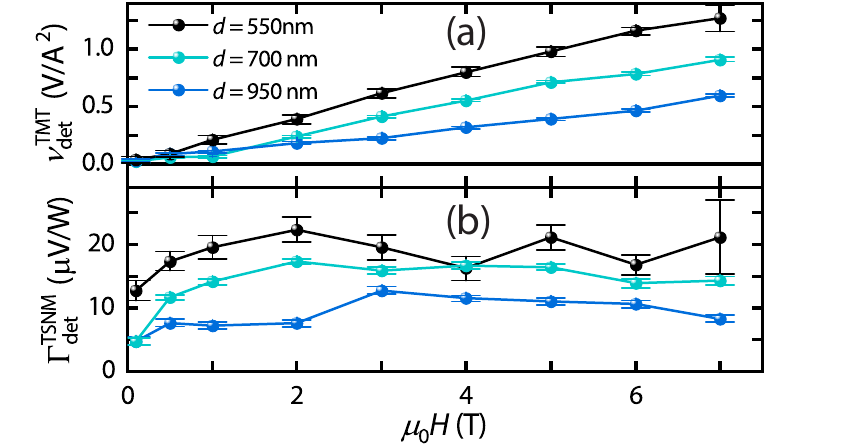}%
	\caption{Signal amplitudes of the TMT (a) and TSNM (b) extracted from the fits of the experimental data shown in Fig.~\ref{fig:ADMR}(a) at $T = \SI{200}{\kelvin}$. (a) TMT signal amplitude $\nu_\mathrm{det}^\mathrm{TMT}$ plotted as a function of the external magnetic field strength for different heater-detector distances $d$. Each of the devices shows a linearly increasing TMT signal for increasing field strength. (b) TSNM signal $\Gamma_\mathrm{det}^\mathrm{TSNM}$ as a function magnetic field, showing an increase of the signal at low field strength and a saturation above $\sim\SI{2}{\tesla}$.
	}%
	\label{fig:Amps}%
\end{figure}

In a next step, we extract the signal amplitudes $\Delta V_\mathrm{det}^\mathrm{TMT}$ and $\Delta V_\mathrm{det}^\mathrm{TSNM}$ from the fits shown in Fig.~\ref{fig:ADMR}(a). To compare the TMT signals between different heater geometries and heater currents $I_\mathrm{q}$, we define the normalized signal amplitudes $\nu_\mathrm{det}^\mathrm{TMT} = ( \Delta V_\mathrm{det}^\mathrm{TMT}/I_\mathrm{q}^2) \cdot (A_\mathrm{h}/A_\mathrm{det})$, where $A_\mathrm{h}$ and $A_\mathrm{det}$ account for the heater and detector areas interfacing the hematite film~\cite{Wimmer2019CoFe}. Regarding the TSNM signals, we normalize the voltage signals to the heater power $P_\mathrm{heat} = R_\mathrm{inj} I_\mathrm{q}^2$ and define $\Gamma_\mathrm{det}^\mathrm{TSNM} = \Delta V_\mathrm{det}^\mathrm{TSNM}/P_\mathrm{heat}$ with $R_\mathrm{h}$ the resistance of the heater. The evolution of these amplitudes with the external magnetic field for different heater-detector distances $d$ is shown in Fig.~\ref{fig:Amps}(a) and (b) for $\nu_\mathrm{det}^\mathrm{TMTM}$ and $ \Gamma_\mathrm{det}^\mathrm{TSNM}$, respectively. The TMT signal linearly increases with the external field strength $\mu_0 H$. This is consistent with the picture given in Fig.~\ref{fig:scheme}(c), since the magnitude $\lvert \bm{M}_\mathrm{net}\rvert = M_\mathrm{net}$ is expected to linearly increase as a function of $\mu_0 H$ as $M_\mathrm{net}\sim\chi_\perp H$ (with $\chi_\perp$ the static magnetic susceptibility of hematite in the easy plane phase). The field dependence of the TSNM signal in Fig.~\ref{fig:Amps}(b) indicates an increase for small fields and a saturation above $\sim\SI{2}{\tesla}$. The observed saturation of the signal is expected for the TSNM since it follows the SMR amplitude in Pt/$\alpha$-Fe$_2$O$_3$~\cite{AltiSMR,Opel2018,Opel2020,Gepraegs2020}. 
Below this magnetic field, hematite exhibits a multi-domain state within the easy-plane~\cite{Opel2020}. 
Both the detector signals corresponding to the TMT and TSNM also show a clear decrease of the signal with increasing distance $d$ from the heater. For the TMT, this is due to the diffusive decay of the thermally excited magnon distribution over distance, leading to a signal decrease with increasing $d$~\cite{CornelissenMMR,Klaui2018}. The decreasing TSNM, on the other hand, stems from the decreasing magnitude of the temperature gradient at the detector position with increasing distance from the heater. 

\begin{figure}[t]%
	\includegraphics[]{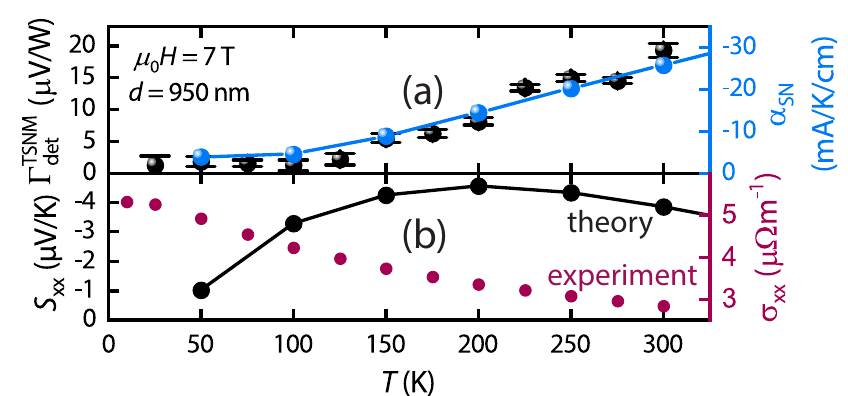}%
	\caption{Normalized TSNM signal amplitude $\Gamma_\mathrm{det}^\mathrm{TSNM}$ extracted from the fits to Eq.~\eqref{eq:fit} plotted as a function of temperature for a device with $d=\SI{950}{\nano\metre}$ at $\mu_0 H = \SI{7}{\tesla}$ (black data points). Blue data points correspond to the theoretical calculation of the spin Nernst conductivity $\alpha_\mathrm{SN}$. (b) Theoretical calculation of the temperature dependence of the conventional (longitudinal) Seebeck coefficient $S_{xx}$ of Pt (black points) and experimentally determined conductivity $\sigma_{xx}$ of the Pt heater (purple points).
	}
	\label{fig:Temp}%
\end{figure}

Finally, we study the temperature dependence of the TSNM signals $\Gamma_\mathrm{det}^\mathrm{TSNM}$ extracted from measurements at $\mu_0 H = \SI{7}{\tesla}$ (black data points in Fig.~\ref{fig:Temp}(a)). We observe a monotonous decrease of the signal with decreasing temperature, which levels out and becomes smaller than the noise level for $T\lesssim\SI{100}{\kelvin}$. 
The relevant scaling parameters to consider for the TSNM are $\Delta V_\mathrm{det}^\mathrm{TSNM} \propto  S_\mathrm{xx} g_\mathrm{r} \lambda_\mathrm{s} \theta_\mathrm{SH}\theta_\mathrm{SN}$~\cite{Meyer2017}, where $S_\mathrm{xx}$ is the longitudinal Seebeck coefficient of Pt, $g_\mathrm{r}$ the real part of the spin mixing interface conductance, $\lambda_\mathrm{s}$ the spin diffusion length of Pt and $\theta_\mathrm{SN}$ as well as $\theta_\mathrm{SH}$ the spin Nernst and spin Hall angle of Pt, respectively. The parameters $g_\mathrm{r}$, $\lambda_\mathrm{s}$ as well as $\theta_\mathrm{SH}$ can be reasonably treated as only weakly temperature dependent~\cite{SibylleSMR,Das2019}. We are thus left with the temperature dependence of both $S_\mathrm{xx}$ and $\theta_\mathrm{SN}$ of Pt. Based on the theoretical description of the SNE in Pt in Ref.~\cite{Meyer2017}, we plot the temperature dependent spin Nernst conductivity $\alpha_\mathrm{SN} = -\theta_\mathrm{SN}\sigma_\mathrm{xx} S_\mathrm{xx}$ (with $\sigma_\mathrm{xx}$ the longitudinal electrical conductivity of Pt) together with our experimental data in Fig.~\ref{fig:Temp}(a) (blue data points). Evidently, the agreement between theory and experiment is excellent, strongly corroborating a decrease of $\alpha_\mathrm{SN}$ with decreasing temperature. In order to determine whether $S_\mathrm{xx}$ or $\theta_\mathrm{SN}$ causes the vanishing TSNM signal, we plot the theoretically calculated $S_\mathrm{xx}$ as a function of temperature in Fig.~\ref{fig:Temp}(b) (black points), showing a finite magnitude even for $T\leq\SI{100}{\kelvin}$. This is also supported by experimental quantifications of $S_\mathrm{xx}$ in bulk Pt~\cite{Moore1973} as well as thin films~\cite{Kockert2019}. Considering that $\theta_\mathrm{SN} = S_\mathrm{yx}^\mathrm{s}/S_\mathrm{xx}$ (with $S_\mathrm{yx}^\mathrm{s}$ the transverse Seebeck coefficient~\cite{Tauber2012}), it follows that $S_\mathrm{yx}^\mathrm{s}$ approaches zero at low temperatures, in accordance with theory~\cite{Tauber2012,SWimmer2013,Popescu_2018}. This is based on the dominance of extrinsic contributions (i.e.~impurity scattering~\cite{SmitSS,BergerSJ}) to the spin Nernst conductivity in Pt. Indeed, the Pt conductivity $\sigma_\mathrm{xx}$ in our sample changes from $\SI{300}{\kelvin}$ to $\SI{10}{\kelvin}$ by about $\sim\SI{45}{\percent}$ (see Fig.~\ref{fig:Temp}(b), purple points), hence exhibiting a finite, non-diverging conductivity at low temperatures. The vanishing TSNM for $T\lesssim\SI{100}{\kelvin}$ can thus be interpreted as a decrease of $\theta_\mathrm{SN}$ towards zero~\footnote{We note, that a quantitative evaluation of the temperature dependence of $\theta_\mathrm{SN}$ is tedious, as it requires the quantification of the temperature gradient present at the detector electrode (which by itself is a temperature dependent quantity) as well as a rigorous experimental evaluation of the temperature dependent Seebeck coefficient $S_\mathrm{xx}$ of thin film Pt.}, an observation which has eluded an experimental observation thus far.

In conclusion, we have investigated the TSNM in a $\alpha$-Fe$_2$O$_3$/Pt device. The excellent agreement of the spin Nernst theory calculations with our data suggests a vanishing spin Nernst angle of Pt at low temperatures and the dominance of extrinsic contributions to the SNE. We demonstrate that the spin Nernst effect is sensitive to the direction of the N\'eel vector of the antiferromagnet, thus representing a suitable platform to discern the TSNM and TMT. Our results shed light on the interaction of pure thermally driven spin currents with antiferromagnets and therefore provide key insights into the physics of pure spin current based magnetothermal effects in AFI/Pt bilayers.

This work is funded by the Deutsche Forschungsgemeinschaft (DFG, German Research Foundation) under Germany’s Excellence Strategy -- EXC-2111 -- 390814868 and project AL2110/2-1.

%

\end{document}